\documentclass[conference]{IEEEtran}
\usepackage{amsfonts}
\usepackage{mathrsfs}
\usepackage{amssymb,amsmath}

\usepackage{algorithm}
\usepackage{algorithmic}
\usepackage{amsmath,amssymb,epsfig,graphics,subfigure}
\usepackage{theorem}
\usepackage{array,color}
\usepackage[compress]{cite}

\theoremheaderfont{\normalfont\bfseries}

\begin{document}

\title{A Unified Linear MSE Minimization MIMO Beamforming Design Based on Quadratic Matrix Programming}
\author{Chengwen Xing$^\dag$, Zesong Fei$^\dag$, Shaodan Ma$^\S$, Jingming Kuang$^\dag$, and Yik-Chung Wu$^\ddag$  \\ $^\dag$School of Information and Electronics, Beijing Institute of Technology, Beijing, China
\\Email: chengwenxing@ieee.org \ \{feizesong, jmkuang\}@bit.edu.cn \\
$^\S$ Department of Electrical and Computer Engineering, University of Macau, Macau\\ Email: shaodanma@umac.mo \\
$^\ddag$ Department of Electrical and Electronic Engineering, The University of Hong Kong, Hong Kong\\
Email: ycwu@eee.hku.hk
 }

\maketitle

\begin{abstract}
In this paper, we investigate a unified linear transceiver design with mean-square-error (MSE) as the objective function for a wide range of wireless systems. The unified design is based on an elegant mathematical programming technology namely quadratic matrix programming (QMP). It is revealed that for different wireless systems such as multi-cell coordination systems, multi-user MIMO systems, MIMO cognitive radio systems, amplify-and-forward MIMO relaying systems, the MSE minimization beamforming design problems can always be solved by solving a number of QMP problems. A comprehensive framework on how to solve QMP problems is also given.
\end{abstract}


\section{Introduction}

In order to satisfy the forecasted data rate boost and to enable high quality multi-media wireless services, more and more available wireless resources are introduced into wireless systems, such as temporal, frequency and spatial resources. The multi-dimensional wireless resources bring new challenges for wireless system designs.
To order realize the promised performance of these resources, the corresponding new technologies will be adopted, such as multiple-carrier technology, multiple-antenna technology and so on.\let\oldthefootnote\thefootnote
\renewcommand\thefootnote{}
\footnote{This research work was supported in part by National Natural Science Foundation of China under Grant No. 61101130 and Sino-Swedish IMT-Advanced and Beyond Cooperative Program under Grant No. 2008DFA11780.}

For spatial resource, multiple-input multiple-output (MIMO) technology is a great success in both theoretical research and practical applications \cite{Telatar1999}. Along with the evolvement of wireless systems, MIMO becomes to be a fundamental and important ingredient of complicated wireless systems e.g., cooperative communication, cognitive communication, physical layer security communication, network coding based communication and so on.
With channel state information (CSI), transceiver designs can significantly improve the performance of MIMO systems \cite{Palomar03}.

Because of a variety of wireless service requirements and wireless environments, there are various wireless systems which have totally different network architectures with even different wireless interfaces. For all of these wireless systems, transceiver design can significantly enhance the whole system performance \cite{Palomar03}. However, for different wireless systems the transceiver design problems admit different signal models, different power constraints and different numbers of variables. It means that these designs should be investigated case by case. From theory research perspective, the theorists want to find a unified design which can reveal some nature of the transceiver designs. To the best our knowledge, the transceiver designs are not unified simultaneously from different performance metrics and different systems.

In the existing works, for unified linear transceiver designs, the widely used logic is for a given system linear transceiver designs with different performance metrics are unified into a general caless of optimization problems with the similar mathematical formulations \cite{Palomar03,Sampth01}. It is well-known that there are two guidelines, i.e., using majorization theory \cite{Palomar03} and weighting operation \cite{Sampth01}. When majorization theory is used, the transceiver design logic is to formulate different performance metrics as different functions of the diagonal elements of the data detection MSE matrix. Furthermore, the functions are classified into two main subclasses of function i.e., Schur-convex or Schur-concave functions. On the other hand, using weighting operation, the different performance metrics are taken into account by using different weighting matrices.

In contrast to these works, in this paper, we give a unified transceiver design which aims at unifying the linear transceiver designs for different wireless systems with the same performance metric minimum mean square error (MMSE). It is revealed that for the beamforming designs in different wireless systems such as multi-cell coordination beamforming design problem, multi-user MIMO beamforming design, cognitive MIMO beamforming design, amplify-and-forward MIMO relaying beamforming design and even their corresponding robust transceiver designs with randomly distributed channel estimation errors and so on, the transceiver design problems can be optimized by solving a series of quadratic matrix programming (QMP) problems. By the way, we also clarify that the QMP problems can always be efficiently solved.

The following notations are used throughout this paper. Boldface
lowercase letters denote vectors, while boldface uppercase letters
denote matrices. The notations  ${\bf{Z}}^{\rm{T}}$, ${\bf{Z}}^{*}$ and ${\bf{Z}}^{\rm{H}}$ denote the transpose, conjugate and
conjugate transpose of the matrix ${\bf{Z}}$, respectively and ${\rm{Tr}}({\bf{Z}})$ is the
trace of the matrix ${\bf{Z}}$. The symbol ${\bf{I}}_{M}$ denotes an
$M \times M$ identity matrix, while ${\bf{0}}_{M,N}$ denotes an $M
\times N$ all zero matrix. The notation ${\bf{Z}}^{1/2}$ is the
Hermitian square root of the positive semi-definite matrix
${\bf{Z}}$, such that ${\bf{Z}}^{1/2}{\bf{Z}}^{1/2}={\bf{Z}}$ and
${\bf{Z}}^{1/2}$ is also a Hermitian matrix. The symbol ${\mathbb{E}}$ denotes statistical expectation operation.

\section{Motivations}

At the beginning, we would like to discuss why our attention is concentrated on linear minimum mean-square-error (LMMSE) transceiver designs in this paper. However for ceratin performance metrics such as bit error rate (BER) the performance of linear transceivers may be not as good as that of their nonlinear counterparts, linear transceivers are still preferred by practical wireless systems due to their low implement complexity.

On the other hand,
MSE is a widely used performance metric of the optimization problems for estimation, detection and transceiver design. It should be pointed out that MSE acting as performance metric suffers from several inherent drawbacks as it is not a terminal performance metric e.g., capacity and BER. Roughly speaking, MSE can be seen as an approximation of the final performance metrics, although they have very closed relationships and particularly, in some special cases, they are even equivalent with each other. Furthermore, MSE is a statistical average, which cannot reflect the detailed information for specific samples.

Maybe more weaknesses can be found, if we discuss in more depth, however MSE is still preferred by the wireless researchers. It may be not the best, but it is not the worst. In general, tractability is the main advantage of MSE. In complicated wireless systems, for several terminal performance metrics, their formulations may be too complicated to prohibit them from application.  From engineers' point of view, the case where there is a solution is much better than that where there is no solution.

In the following, we show an \textbf{example} to illustrate that QMP problems are of great importance in LMMSE transceiver designs. Note that this example is discussed in detail in our previous work \cite{MaMILCOM2010}. Here, it only  provides a prologue of our work in this paper. First, we want to  highlight that the algorithm discussed in this paper is not limited to the example, which has a much wider application range. For example, the algorithm discussed in this paper can also be applied to multi-hop networks.
Moreover, different from our previous work \cite{MaMILCOM2010}, in this paper we provide a comprehensive framework for LMMSE transceiver design. We are not limited to a specific communication systems. We try to reveal the nature of LMMSE transceiver design and answer the questions why QMP should be chosen and how to solve the optimization problem using QMP.

\subsection{An example:}
In the example, a dual-hop amplify-and-forward (AF) relay network is considered, in which  there are multiple source
nodes, relay nodes and destination nodes. Furthermore, different
source can have different numbers of transmit antennas and data
stream to transmit. It is denoted that the number of transmit
antennas of the $i^{\rm{th}}$ source is $N_{S,i}$. It is also assumed that for each source node there may be
more than one corresponding destination node. There are also
multiple relay nodes in the network, and the $j^{\rm{th}}$ relay has
$M_{R,j}$ receive antennas and $N_{R,j}$ transmit antennas. At the first hop, the source nodes transmit
data to the relay nodes. The received signal, ${\bf{x}}_j$, at the
$j^{\rm{th}}$ relay node is
\begin{align}
{\bf{x}}_{j}&={\bf{H}}_{sr,ij}{\sum}_k({\bf{P}}_{ik}{\bf{s}}_{ik})+{\sum}_{l \ne i}[{\bf{H}}_{sr,lj}{\sum}_k({\bf{P}}_{lk}{\bf{s}}_{lk})]\nonumber \\
&+{\bf{n}}_{1,j}.
\end{align}
where ${\bf{s}}_{ik}$ is the data vector transmitted by the
$i^{\rm{th}}$ source node to the $k^{\rm{th}}$ destination with the
covariance matrix
${\bf{R}}_{{\bf{s}}_{ik}}=\mathbb{E}\{{\bf{s}}_{ik}{\bf{s}}_{ik}^{\rm{H}}\}$.
When the $i^{\rm{th}}$ source node does not want to transmit signal
to $k^{\rm{th}}$ destination, ${\bf{s}}_{ik}$ is a all-zero vector.

At the source, before transmission the signal is multiplied a
precoder ${\bf{P}}_{ik}$ under transmit power constraint
$\sum_k{\rm{Tr}}({\bf{P}}_{ik}{\bf{R}}_{{\bf{s}}_{ik}}{\bf{P}}_{ik}^{\rm{H}})\le
P_{s,i}$, where $P_{s,i}$ is the maximum transmit power at the
$i^{\rm{th}}$ source node. The matrix ${\bf{H}}_{sr,ij}$ is the MIMO
channel matrix between the $i^{\rm{th}}$ source node and the
$j^{\rm{th}}$ relay node. Symbol ${\bf{n}}_{1,j}$ is the additive
Gaussian noise with the covariance matrix
${\bf{R}}_{{\bf{n}}_{1,j}}$. At the $j^{\rm{th}}$ relay node, the
received signal ${\bf{x}}_{j}$ is multiplied by a precoder matrix
${\bf{F}}_j$, under a power constraint
${\rm{Tr}}({\bf{F}}_j{\bf{R}}_{{\bf{x}}_j}{\bf{F}}_j^{\rm{H}}) \le
P_{r,j}$ where
${\bf{R}}_{{\bf{x}}_j}=\mathbb{E}\{{\bf{x}}_j{\bf{x}}_j^{\rm{H}}\}$
and $P_{r,j}$ is the maximum transmit power. Then the resulting
signal is transmitted to the destination. The received signal at the
$k^{\rm{th}}$ destination, ${\bf{y}}_k$, can be written as
\begin{align}
\label{signal_destin}
{\bf{y}}_{k} & ={\sum}_{j}({\bf{H}}_{rd,jk}{\bf{F}}_{j}{\bf{x}}_{j})+{\bf{n}}_{2,i} \nonumber \\
&={\sum}_{j}[{\bf{H}}_{rd,jk}{\bf{F}}_j{\sum}_l({\bf{H}}_{sr,lj}{\bf{P}}_{lk}{\bf{s}}_{lk})]
\nonumber \\
&+{\sum}_j[{\bf{H}}_{rd,ji}{\bf{F}}_j{\sum}_l({\bf{H}}_{sr,lj}{\sum}_{m \ne k }({\bf{P}}_{lm}{\bf{s}}_{lm}))]\nonumber \\
&+{\sum}_j({\bf{H}}_{rd,jk}{\bf{F}}_j{\bf{n}}_{1,j}) +{\bf{n}}_{2,k}.
\end{align} where ${\bf{H}}_{rd,jk}$ is the MIMO channel matrix between the $j^{\rm{th}}$ relay
and the $k^{\rm{th}}$ destination, and ${\bf{n}}_{2,k}$ is the additive Gaussian noise
vector at the second hop with covariance matrix ${\bf{R}}_{{\bf{n}}_{2,k}}$.

The optimization problem of linear minimum mean-square-error (LMMSE) transceiver design can be formulated as \cite{MaMILCOM2010}
\begin{align}
\label{prob:opt}
& \min \ \ \ {\sum}_k {\rm{MSE}}_k={\mathbb{E}}\{\|{\bf{G}}_k{\bf{y}}_k-[{\bf{s}}_{1k}^{\rm{T}},\cdots,{\bf{s}}_{N_sk}^{\rm{T}}]^{\rm{T}}\|^2\} \nonumber \\
& \ {\rm{s.t.}} \ \ \ \ {\rm{Tr}}({\bf{F}}_j{\bf{R}}_{{\bf{x}}_j}{\bf{F}}_j^{\rm{H}})\le P_{r,j} \ \ j\in {\mathcal{E}}_{r} \nonumber \\
& \ \ \ \ \ \ \ \ \ {\sum}_k {\rm{Tr}}({\bf{P}}_{ik}{\bf{R}}_{{\bf{s}}_{ik}}{\bf{P}}_{ik}^{\rm{H}} )\le P_{s,i} \ \ i\in {\mathcal{E}}_{s}
\end{align}where $[{\bf{s}}_{1k}^{\rm{T}},\cdots,{\bf{s}}_{N_sk}^{\rm{T}}]^{\rm{T}}$ is the desired signal to be recovered at the $k^{\rm{th}}$ destination, and ${\mathcal{E}}_{r} $ and ${\mathcal{E}}_{s}$ denote the set of relay nodes and the set of source nodes, respectively.

The optimization problem (\ref{prob:opt}) is a very general problem which includes the following scenarios as its special cases.

\noindent $\bullet$  Multi-user MIMO uplink transceiver design  \cite{XingSP2012}.

\noindent $\bullet$  Multi-user MIMO downlink transceiver design \cite{XingSP2012}.

\noindent $\bullet$ Two-way AF MIMO relaying LMMSE transceiver design \cite{XingICSPCC2011}. Two-way AF MIMO relay can be taken as a soft combination of uplink and downlink beamforming designs.

\noindent $\bullet$ Multi-cell coordination beamforming design.

As there are too many variables to be optimized and the nonconvex nature of the optimization problem makes the problem more complicated, it is difficult to find a closed-form globally optimal solution. In order to carry out the transceiver designs, several suboptimal solutions are usually proposed. Iterative algorithm is one of the most widely used and important suboptimal solutions.
We admit that iterative algorithms suffer from some well-known weaknesses. First, the final solution is greatly affected by the initial value selection. Second, the convergence of an iterative algorithm should be guaranteed. If not, the iterative algorithm may be meaningless.
Third, in general even with proved convergence there is no guarantee that the final solution is globally optimal. However, iterative algorithms still have two important characteristics making them preferable. First, it can be applied to a much wide area of transceiver designs ranging from traditional a point-to-point system to a distributed network. Second, it can act as a performance benchmark for other suboptimal solutions. Inspired by these, we focus on iterative algorithms in this paper.

\subsection{Quadratic nature of the LMMSE transceiver designs}

The MSE is an integration over the signals and noises. From its name, it is direct and obvious that MSE is a  quadratic formulation. Moreover, in this paper, we concentrate our attention to the case where the variables are matrices, as in MIMO systems the variables to be optimized are always matrices. Inspired by these facts, a kind of function termed as quadratic matrix (QM) functions with a complex matrix variable ${\bf{X}}$ is defined as
\begin{align}
f_l({\bf{X}})={\rm{Tr}}({\bf{D}}_l{\bf{X}}^{\rm{H}}{\bf{A}}_l{\bf{X}})+2{\mathcal R}\{{\rm{Tr}}({\bf{B}}_l^{\rm{H}}{\bf{X}})\}+c_l
\end{align}where ${\bf{A}}_l={\bf{A}}_l^{\rm{H}} \in {\mathbb C}^{n\times n}$, ${\bf{B}}_l\in {\mathbb C}^{n\times r}$, $c_l \in {\mathbb R}$, ${\bf{D}}_l={\bf{D}}_l^{\rm{H}} \in {\mathbb C}^{r\times r}$. In addition, ${\mathcal R}\{\bullet\}$ denotes the real part. It can be seen that a QM function consists of three terms which are second-order term, first-order term and zeroth-order term. If the following conditions are satisfied, no matter what the system is, the MSE with linear transceiver is a QM function with respective to each variable, separately.

\noindent (1). The considered system is a linear system. The linearity is mainly referred to the following two properties:

(a.1) The received signal at the destination is a linear function of the transmit signal when all variables are fixed.

(a.2) The received signal at the destination is a linear function with respective to each variable when the signal is fixed and the other variables are fixed.

\noindent (2). The desired signals are independent of the noises.
 It means that when the signal vector  is denoted by ${\bf{s}}$ and the equivalent noise vector is ${\bf{v}}$, the following equality must hold
 \begin{align}
\mathbb{E}\{ {\bf{s}}{\bf{v}}^{\rm{H}}\}={\bf{0}}.
 \end{align}

In addition, the constraints in transceiver design for wireless systems are usually QM functions, as the constraints are usually related with energy, which are definitely  quadratic terms, e.g., transmit power, interference to primary users, and so on. Therefore, most of LMMSE transceiver designs can be iteratively optimized by solving a number of optimization problems consisting of QM functions in both objective function and constraint functions. This kind of optimization problem is named as quadratic matrix programming (QMP) problems. In this following, the properties of QMP will be discussed in detail as well as how to solve it.

\section{Fundamentals of QMP}
\label{Sect:QMP}
At the beginning of this section, the definition and properties of QMP are investigated.
Although in \cite{Beck07}, a definition of quadratic matrix programming is given, in this paper we first revise the definition given in \cite{Beck07} in order to accommodate more cases.  Our definition is more general and has a wider range of applications. A standard QMP problem is defined as
\begin{align}
\label{optimization_P_1}
& \textbf{Type 1 QMP:} \nonumber \\
& \min_{\bf{X}} \ \ {\rm{Tr}}({\bf{D}}_0{\bf{X}}^{\rm{H}}{\bf{A}}_0{\bf{X}})+2{\mathcal R}\{{\rm{Tr}}({\bf{B}}_0^{\rm{H}}{\bf{X}})\}+c_0 \nonumber \\
& \ {\rm{s.t.}} \ \ \ {\rm{Tr}}({\bf{D}}_i{\bf{X}}^{\rm{H}}{\bf{A}}_i{\bf{X}})+2{\mathcal R}\{{\rm{Tr}}({\bf{B}}_i^{\rm{H}}{\bf{X}})\}+c_i \le 0, i \in {\mathcal I} \nonumber \\
& \ \ \ \ \ \ \ \ {\rm{Tr}}({\bf{D}}_j{\bf{X}}^{\rm{H}}{\bf{A}}_j{\bf{X}})+2{\mathcal R}\{{\rm{Tr}}({\bf{B}}_j^{\rm{H}}{\bf{X}})\}+c_j=0, j \in {\mathcal E} \nonumber \\
& \ \ \ \ \ \ \ \ {\bf{X}} \in {\mathbb C}^{n\times r}
\end{align}where $l \in \{0\} \cup {\mathcal{I}} \cup {\mathcal {E}}$. These assumptions are essential to guarantee that the objective function and constraint functions are real-valued functions, as it is meaningless to minimize a complex-valued function. It is obvious that QMP is a special case of quadratically constrained quadratic programming (QCQP), which is a class of very famous and widely used optimization problems. Then it is natural that QMP has much better properties than QCQP, which can be exploited to solve the involved optimization problems. This is exactly the motivation of the research on QMP \cite{Beck07,Beck09}.

\noindent \textbf{General QMP:}

Based on the
properties of Kronecker product and the following definitions
\begin{align}
{\boldsymbol \Omega}_l \triangleq \left[ {\begin{array}{*{20}c}
  {\bf{D}}_{l}^{\rm{T}} \otimes {\bf{A}}_l & {\rm{vec}}({{\bf{B}}_l)}  \\
   {\rm{vec}}^{\rm{H}}({\bf{B}}_l) & {c_l}   \\
\end{array}} \right],
\end{align}the optimization problem (\ref{optimization_P_1}) is equivalent to
\begin{align}
\label{SDP}
& {\min} \ \ \ {\rm{Tr}}({\boldsymbol \Omega}_0{\bf{Z}}) \nonumber \\
& {\rm{s.t.}} \ \ \ \ {\rm{Tr}}({\boldsymbol \Omega}_i{\bf{Z}}) \le
0, \ \ {\rm{Tr}}({\boldsymbol \Omega}_j{\bf{Z}})=0
\nonumber \\
&\ \ \ \ \ \ \ \  {\bf{Z}}=[{\rm{vec}}^{\rm{T}}({\bf{X}}) \
1]^{\rm{T}}[{\rm{vec}}^{\rm{H}}({\bf{X}}) \ 1].
\end{align}If the constraint ${\rm{Rank}}({\bf{Z}})=1$ is relaxed
(it is a well-known semi-definite relaxation (SDR) \cite{Beck09}), we have the following semi-definite programming (SDP) problem \cite{Vandenberghe96},  which can be efficiently solved by
interior point polynomial algorithms
\begin{align}
\label{SDR}
& {\min\limits_{\bf{Z}}} \ \ \ {\rm{Tr}}({\boldsymbol \Omega}_0{\bf{Z}}) \nonumber \\
& {\rm{s.t.}} \ \ \ \ {\rm{Tr}}({\boldsymbol \Omega}_i{\bf{Z}}) \le
0, \ \  {\rm{Tr}}({\boldsymbol \Omega}_j{\bf{Z}})=0
\nonumber \\
&\ \ \ \ \ \ \ \  [{\bf{Z}}]_{NN_s+1,NN_s+1}=1, \ \ {\bf{Z}}
\succeq 0,
\end{align}where ${\bf{Z}}$ is a Hermitian matrix. When ${\bf{A}}_l$ and ${\bf{D}}_l$ are both positive semi-definite matrices and there are only inequality constraints, the optimization problem (\ref{optimization_P_1}) is a convex optimization problem. For convex optimization problem, we do not need the previous SDR to solve the optimal solutions. In the following, two methods for the convex case are proposed.

\noindent \textbf{Convex QMP:}

When ${\bf{A}}_l$ and ${\bf{D}}_l$ are both positive semi-definite matrices with inequality constraints, the constraints only exists inequality constraints.

\noindent \underline{\textbf{SDP Based Algorithm:}}

Using the properties of Kronecker product ${\rm{Tr}}({\boldsymbol{A}}{\boldsymbol{B}})={\rm{vec}}^{\rm{H}}({\boldsymbol{A}}^{\rm{H}}){\rm{vec}}({\boldsymbol{B}})$, the QM function can be reformulated as
\begin{align} &{\rm{Tr}}({\bf{D}}_l^{\rm{H}/2}{\bf{X}}^{\rm{H}}{\bf{A}}_l
{\bf{X}}{\bf{D}}_l^{1/2})
+2\mathcal{R}\{{\rm{Tr}}({\bf{B}}_l^{\rm{H}}{\bf{X}})\}+c_l \nonumber \\
&={\rm{Tr}}({\bf{D}}_l^{\rm{H}/2}{\bf{X}}^{\rm{H}}{\bf{A}}_l^
{\frac{\rm{H}}{2}}
{\bf{A}}_l^{\frac{1}{2}}
{\bf{X}}{\bf{D}}_l^{1/2})
+2\mathcal{R}\{{\rm{Tr}}({\bf{B}}_l^{\rm{H}}{\bf{X}})\}+c_l \nonumber \\
&={\rm{vec}}^{\rm{H}}({\bf{X}})({\bf{D}}_l^{*/2}\otimes {\bf{A}}_l^{\frac{\rm{H}}{2}})({\bf{D}}_l^{\rm{T}/2}\otimes {\bf{A}}_l^{\frac{1}{2}}){\rm{vec}}({\bf{X}})\nonumber \\
&+2\mathcal{R}\{{\rm{vec}}^{\rm{H}}({\bf{B}}_l){\rm{vec}}({\bf{X}})\}+c_l
\le 0,
\end{align} based on which and together with Schur complement lemma, the optimization problem (\ref{prob:opt}) can be reformulated as the SDP problem given at the top of the next page.
\begin{figure*}[!t]
\normalsize
\begin{align}
\label{SDR}
&{\min} \ \ \ \ \ \ \ \  t \nonumber \\
& \ {\rm{s.t.}}  \ \
 \left[ {\begin{array}{*{20}c}
   {\bf{I}} & ({\bf{D}}_0^{\frac{\rm{T}}{2}}\otimes {\bf{A}}_0^{\frac{1}{2}}){\rm{vec}}({\bf{X}})   \\
   (({\bf{D}}_0^{\frac{\rm{T}}{2}}\otimes {\bf{A}}_0^{\frac{1}{2}}){\rm{vec}}({\bf{X}}))^{\rm{H}} & -2\mathcal{R}({\rm{vec}}^{\rm{H}}({\bf{B}}_0){\rm{vec}}({\bf{X}}))
   +t  \\
\end{array}} \right]  \succeq 0 \nonumber  \\
& \ \ \ \ \ \ \ \left[ {\begin{array}{*{20}c}
   {\bf{I}} & ({\bf{D}}_i^{\frac{\rm{T}}{2}}\otimes {\bf{A}}_i^{\frac{1}{2}}){\rm{vec}}({\bf{X}})   \\
   (({\bf{D}}_i^{\frac{\rm{T}}{2}} \otimes {\bf{A}}_i^{\frac{1}{2}}){\rm{vec}}({\bf{X}}))^{\rm{H}} & -2\mathcal{R}({\rm{vec}}^{\rm{H}}({\bf{B}}_i){\rm{vec}}
   ({\bf{X}}))-c_i \\
\end{array}} \right] \succeq 0
\end{align}
\hrulefill
\end{figure*}

Notice that in our work, the variables are complex matrices. For some optimization tool boxes, only real variables are permitted. In that case, a minor transformation is needed, which is
\begin{align}
\left[ {\begin{array}{*{20}c}
  {\bf{I}}_N &  {\bf{v}} \\
    {\bf{v}}^{\rm{H}}& a  \\
\end{array}} \right] \succeq 0 \rightarrow \left[ {\begin{array}{*{20}c}
  {\bf{I}}_{2N} &  {\bf{\tilde v}} \\
    {\bf{\tilde v}}^{\rm{T}}& a  \\
\end{array}} \right] \succeq 0
\end{align} where $ {\bf{\tilde v}}$ is defined as
\begin{align}
 {\bf{\tilde v}}=[{\rm{Real}}({\bf{ v}})^{\rm{T}} \ {\rm{Imag}}({\bf{ v}})^{\rm{T}}]^{\rm{T}}.
\end{align}

When ${\bf{A}}_i$ and ${\bf{D}}_i$ are positive definite matrices, the optimization problem can be further transformed into a more efficient convex optimization problem termed as second order conic programming (SOCP) problems.

\noindent \underline{\textbf{SOCP Based Algorithm:}}

Notice the QM functions in both the objective function and constraints can be reformulated as
\begin{align}
&{\rm{Tr}}({\bf{D}}_l^{\rm{H}/2}{\bf{X}}^{\rm{H}}{\bf{A}}_l
{\bf{X}}{\bf{D}}_l^{1/2})
+2\mathcal{R}\{{\rm{Tr}}({\bf{B}}_l^{\rm{H}}{\bf{X}})\}+c_l\nonumber \\
=&\left\| \left[{\begin{array}{*{20}c}
   {{\bf{A}}_l^{\frac{1}{2}}{\bf{X}}{\bf{D}}_l^{\frac{1}{2}}
   +{\bf{A}}_l^{-\frac{1}{2}}{\bf{B}}_l{\bf{D}}_l^{-\frac{1}{2}}}  \\
\end{array}}\right] \right\|_{\rm{F}}^2+c_i\nonumber \\
&-{\rm{Tr}}({\bf{A}}_l^{-1}{\bf{B}}_l{\bf{D}}_l^{-1}{\bf{B}}_l^{\rm{H}})
\end{align}where $\|\bullet\|_{\rm{F}}$ denotes the Frobenius norm. Therefore,
the optimization problem (\ref{optimization_P_1}) can be reformulated as a standard
 SOCP problem which is given at the top of the next page.

\begin{figure*}[!t]
\normalsize
\begin{align}
& \min_{{\bf{P}}_k,t} \ \ \ t \nonumber \\
& {\rm{s.t.}} \ \ \ \left\| \left[{\begin{array}{*{20}c}
   {{\bf{A}}_0^{\frac{1}{2}}{\bf{X}}{\bf{D}}_0^{\frac{1}{2}}
   +{\bf{A}}_0^{-\frac{1}{2}}{\bf{B}}_0{\bf{D}}_0^{-\frac{1}{2}}}  \\
\end{array}}\right] \right\|_{\rm{F}}\le t \nonumber \\
& \ \ \ \ \ \ \ \left\| \left[{\begin{array}{*{20}c}
   {{\bf{A}}_i^{\frac{1}{2}}{\bf{X}}{\bf{D}}_i^{\frac{1}{2}}
   +{\bf{A}}_i^{-\frac{1}{2}}{\bf{B}}_i{\bf{D}}_i^{-\frac{1}{2}}}  \\
\end{array}}\right]  \right\|_{\rm{F}} \le \sqrt{{\rm{Tr}}({\bf{A}}_i^{-1}{\bf{B}}_i{\bf{D}}_i^{-1}{\bf{B}}_i^{\rm{H}})-c_i}
\end{align}
\hrulefill
\end{figure*}

In the remaining part of this paper, we will concentrate our attention to a special QMP defined as
\begin{align}
\label{optimization_P_2}
& \textbf{Type 2 QMP:} \nonumber \\
& \min_{\bf{X}} \ \ {\rm{Tr}}({\bf{X}}^{\rm{H}}{\bf{A}}_0{\bf{X}})+2{\mathcal R}\{{\rm{Tr}}({\bf{B}}_0^{\rm{H}}{\bf{X}})\}+c_0 \nonumber \\
& \ {\rm{s.t.}} \ \ \ {\rm{Tr}}({\bf{X}}^{\rm{H}}{\bf{A}}_i{\bf{X}})+2{\mathcal R}\{{\rm{Tr}}({\bf{B}}_i^{\rm{H}}{\bf{X}})\}+c_i \le 0, i \in {\mathcal I} \nonumber \\
& \ \ \ \ \ \ \ \ {\rm{Tr}}({\bf{X}}^{\rm{H}}{\bf{A}}_j{\bf{X}})+2{\mathcal R}\{{\rm{Tr}}({\bf{B}}_j^{\rm{H}}{\bf{X}})\}+c_j=0, j \in {\mathcal E} \nonumber \\
& \ \ \ \ \ \ \ \ {\bf{X}} \in {\mathbb C}^{n\times r}
\end{align}which has much better properties than the general Type-1 QMP \cite{Beck07}.
For notational simplicity, in the remaining part of the paper, the T-2-QMP problems are referred to as the Type 2 QMP problem.

\section{Properties of T-2-QMP}
\subsection{T-2-QMP without Constraints}
At the first glance, we discuss the case without constraint which reads as
\begin{align}
\label{Opt_without}
& \min_{\bf{X}} \ \ {\rm{Tr}}({\bf{X}}^{\rm{H}}{\bf{A}}_0{\bf{X}})+2{\mathcal R}\{{\rm{Tr}}({\bf{B}}_0^{\rm{H}}{\bf{X}})\}+c_0
\end{align}where ${\bf{A}}_0>{\bf{0}}$. This case corresponds to linear minimum mean square error (LMMSE) equalizer design, which is also named as LMMSE estimator design. It is obvious that the optimization problem (\ref{Opt_without}) is convex and then the optimal solution is exactly the solution satisfying its differentiation equals to 0, i.e., $
{\bf{A}}_0{\bf{X}}=-{\bf{B}}_0 $. Specifically, the optimal solution has the following closed-form solution
\begin{align}
{\bf{X}}_{\rm{opt}}=-{\bf{A}}_0^{-1}{\bf{B}}_0.
\end{align}

Moreover, considering weighted MSE minimization, the optimization problem becomes to be
\begin{align}
\label{weighted MSE optimal solution}
& {\min}_{\bf{X}} \ \ {\rm{Tr}}({\bf{W}}{\bf{X}}^{\rm{H}}{\bf{A}}_0{\bf{X}})+2{\mathcal R}\{{\rm{Tr}}({\bf{W}}^{\rm{H}}{\bf{B}}_0^{\rm{H}}{\bf{X}})\}+c_0
\end{align}where ${\bf{W}}\succeq {\bf{0}}$ is a weighting matrix. Following the same logic as previously discussed, we have the optimal solution must satisfy
\begin{align}
\label{opt_conditions}
{\bf{A}}_0{\bf{X}}{\bf{W}}=-{\bf{B}}_0{\bf{W}}
\end{align}Because ${\bf{W}}$ can be ill-rank, the optimal solution is not unique. It is obvious that ${\bf{X}}_{\rm{opt}}=-{\bf{A}}_0^{-1}{\bf{B}}_0$ satisfying the previous condition, which is exactly the optimal solution.

\noindent \underline{\textbf{Conclusion 1:}} Without constraints, the optimal solution ${\bf{X}}_{\rm{opt}}$ has a closed form. Notice that ${\bf{X}}_{\rm{opt}}^{\rm{H}}$ is just the Wiener filter. It is well-known for a linear system with Gaussian noise, LMMSE equalizer is exactly the optimal equalizer \cite{Kay93}.

\subsection{T-2-QMP with One Constraint}

In this section, we focus on the case where there is only one constraint in a QMP problem. This case corresponds to the scenario there is only one transmit power constraint. Here  we focus on the following T-2-QMP problem
\begin{align}
\label{Prob_one}
& \min_{\bf{X}} \ \ {\rm{Tr}}({\bf{X}}^{\rm{H}}{\bf{A}}_0{\bf{X}})+2{\mathcal R}\{{\rm{Tr}}({\bf{B}}_0{\bf{X}})\}+c_0 \nonumber \\
& \ {\rm{s.t.}} \ \ \ {\rm{Tr}}({\bf{X}}^{\rm{H}}{\bf{A}}_1{\bf{X}}) \le P,
\end{align}where ${\bf{A}}_1>{\bf{0}}$. Without loss of generality, it is assumed that the feasible set of the considered optimization problem is not empty. In this scenario, solving an unknown matrix variable can be reduced to solving an unknown scalar variable. Therefore, the computation dimensionality and complexity are significantly reduced. In this following, we will clarify this step by step.

For constrained optimization problems, when regularity conditions are satisfied, Karush-Kuhn-Tucker (KKT) are the necessary conditions for the optimal solutions and then they can provide useful information to derive the optimal solutions. For the optimization problem (\ref{Prob_one}) with one constraint, the regularity condition termed as linear independence of constraint qualification (LICQ) can be easily proved to hold and thus KKT conditions are the necessary conditions. The corresponding Lagrange function of the optimization problem (\ref{Prob_one}) is expressed as
\begin{align}
\label{Lagrange_function}
\mathcal{L}({\bf{X}})&={\rm{Tr}}({\bf{X}}^{\rm{H}}{\bf{A}}_0{\bf{X}})+2{\mathcal R}\{{\rm{Tr}}({\bf{B}}_0^{\rm{H}}{\bf{X}})\}+c_0\nonumber \\
& +\mu( {\rm{Tr}}({\bf{X}}^{\rm{H}}{\bf{A}}_1{\bf{X}})-P),
\end{align}where $\mu\ge0$ is the Lagrange multiplier. Based on (\ref{Lagrange_function}),  the KKT conditions of the optimization problem (\ref{Prob_one}) can be derived to be \cite{Boyd04}
\begin{align}
&({\bf{A}}_0+\mu{\bf{A}}_1){\bf{X}}=-{\bf{B}}_0 \\
&\mu( {\rm{Tr}}({\bf{X}}^{\rm{H}}{\bf{A}}_1{\bf{X}})-P)=0  \\
&{\rm{Tr}}({\bf{X}}^{\rm{H}}{\bf{A}}_1{\bf{X}}) \le P \\
& \mu\ge0
\end{align}

In this case, the optimal solution has the following semi-closed-form solution
\begin{align}
\label{X_solution}
{\bf{X}}=-({\bf{A}}_0+\mu{\bf{A}}_1)^{-1}{\bf{B}}_0
\end{align}in which the only unknown variable is a scalar Lagrange multiplier. Substituting (\ref{X_solution}) into the constraint of (\ref{Prob_one}), we have
\begin{align}
&{\rm{Tr}}({\bf{X}}^{\rm{H}}{\bf{A}}_1{\bf{X}})\nonumber \\
=&{\rm{Tr}}
({\bf{B}}_0^{\rm{H}}({\bf{A}}_0+\mu{\bf{A}}_1)^{-1}{\bf{A}}_1({\bf{A}}_0+\mu{\bf{A}}_1)^{-1}{\bf{B}}_0)\nonumber \\
=&{\rm{Tr}}
({\bf{B}}_0^{\rm{H}}{\bf{A}}_1^{-\frac{1}{2}}({\bf{A}}_1^{-\frac{1}{2}}
{\bf{A}}_0{\bf{A}}_1^{-\frac{1}{2}}+\mu{\bf{I}})^{-2}{\bf{A}}_1^{-\frac{1}{2}}{\bf{B}}_0) \nonumber \\
\triangleq & g(\mu).
\end{align}In Appendix A, it has been proved that $g(\mu)$ is a decreasing function with respective to $\mu$, and its value can be computed by using a simple one-dimensional search such as bisection search. Based on this conclusion and the KKT conditions given previously, the value of $\mu$ can be computed to be
\begin{equation}
\label{gamma}
\mu=\begin{cases} 0 & \text{if $g(0) \le P$} \\
\text{Solve $g(\mu)=P$ } &
\text{Otherwise}
\end{cases}.
\end{equation}

It can be seen that the solution satisfying KKT conditions is unique.
 Notice that the KKT conditions are the necessary conditions for the optimal solutions. As a result, the unique solution satisfying the KKT conditions is exactly the optimal solution.
This is of great importance, because the unknown variable is simplified from a matrix to a scalar. In other words, the number of variables is significantly reduced and the corresponding computational complexity is also significantly reduced.



\noindent \underline{\textbf{Conclusion 2:}} With only one constraint, the T-2-QMP problem has a semi-closed-form solution with an unknown scalar variable. The unknown scalar variable can be efficiently computed using one dimensional search.

\noindent \underline{\textbf{Remark:}} Notice that in Boyd's book \cite{Boyd04}, it never states that the KKT conditions are the necessary conditions for the optimal solutions without any prior conditions. It is possible that KKT conditions are not the necessary conditions.

\subsection{T-2-QMP with more than one constraint}
For T-2-QMP problems with more than one constraint, solving the optimization problems must also rely on numerical algorithms such as interior point algorithms. As a T-2-QMP problems has much better structures compared to the general QMP problem discussed in Section~\ref{Sect:QMP}, it exhibits  stronger convexity property which can be exploit to efficiently solve the optimization problem.
As discussed in \cite{Beck09}, the original optimization problem is first transformed into its homogenized problem which can be efficiently solved. First, the homogenized QM function of the QM function defined previously is denoted by $f_i^{\rm{H}}$
\begin{align}
f_l^{\rm{H}}({\bf{Y}};{\bf{Z}})=&{\rm{Tr}}({\bf{Y}}^{\rm{H}}{\bf{A}}_l{\bf{Y}})+2{\mathcal R}\{{\rm{Tr}}({\bf{Z}}^{\rm{H}}{\bf{B}}_l^{\rm{H}}{\bf{Y}})\}\nonumber \\
&+{c_l}/{r}{\rm{Tr}}({\bf{Z}}^{\rm{H}}{\bf{Z}}).
\end{align}Then introducing the following operators,
\begin{align}
{\bf{M}}_l(f_l)=
\left[ {\begin{array}{*{20}c}
   {{\bf{A}}_l} & {{\bf{B}}_l}  \\
   {{\bf{B}}_l^{\rm{H}}} & {\frac{c_l}{r}{\bf{I}}_r}  \\
\end{array}} \right]
\end{align} the homogenized optimization problem of (\ref{optimization_P_2}) is formulated as
\begin{align}
& \min \ \ {\rm{Tr}}({\bf{M}}(f_0)[{\bf{Y}};{\bf{Z}}][{\bf{Y}};{\bf{Z}}]^{\rm{H}}) \nonumber \\
& \ {\rm{s.t.}} \ \ \ {\rm{Tr}}({\bf{M}}(f_i)[{\bf{Y}};{\bf{Z}}][{\bf{Y}};{\bf{Z}}]^{\rm{H}}) \le \alpha_i, i \in {\mathcal I} \nonumber \\
& \ \ \ \ \ \ \ \ {\rm{Tr}}({\bf{M}}(f_j)[{\bf{Y}};{\bf{Z}}][{\bf{Y}};{\bf{Z}}]^{\rm{H}})=\alpha_j, j \in {\mathcal E} \nonumber \\
& \ \ \ \ \ \ \ \ {\bf{Z}}^{\rm{H}}{\bf{Z}}={\bf{I}}_r  \ \ \ \ {\bf{Y}}\in {\mathbb C}^{n\times r}.
\end{align}Notice that the optimal solution of (\ref{optimization_P_2}) ${\bf{X}}_{\rm{opt}}$ equals to
\begin{align}
{\bf{X}}_{\rm{opt}}={\bf{Y}}_{\rm{opt}}{\bf{Z}}_{\rm{opt}}^{\rm{H}}.
\end{align}
Defining ${\bf{U}}\triangleq [{\bf{Y}};{\bf{Z}}][{\bf{Y}};{\bf{Z}}]^{\rm{H}}$, after relaxing the rank constraint on ${\bf{U}}$, we have the following optimization problem
\begin{align}
& \min_{\bf{U}} \ \ {\rm{Tr}}({\bf{M}}(f_0){\bf{U}}) \nonumber \\
& \ {\rm{s.t.}} \ \ \ {\rm{Tr}}({\bf{M}}(f_i){\bf{U}}) \le \alpha_i, i \in {\mathcal I} \nonumber \\
& \ \ \ \ \ \ \ \ {\rm{Tr}}({\bf{M}}(f_j){\bf{U}})=\alpha_j, j \in {\mathcal E} \nonumber \\
& \ \ \ \ \ \ \ \ [{\bf{U}}]_{n+1:n+r,n+1:n+r}={\bf{I}}_r \ \ \ \ {\bf{U}}\succeq {\bf{0}}.
\end{align}
To recover ${\bf{X}}$ from ${\bf{U}}$, a rank reduction based algorithm has been given in \cite{Beck07}. When the number of the constraints are less than $2r$, this relaxation is tight \cite{Beck09}. It is concluded that T-2-QMP has a much stronger convexity property than the general QMP.

\noindent \underline{\textbf{Remark:}} From the practical viewpoint, due to the limited length of training sequences and the time varying nature of wireless channels, channel estimation errors are always inevitable. When channel errors are taken into account, the channel state information can be written as ${\bf{H}}_{l}={\bf{\hat H}}_{l}+\Delta{\bf{H}}_{l}$ where ${\bf{\hat H}}_{l}$ is the estimated ${\bf{H}}_{l}$ and $\Delta{\bf{H}}_{l}$ is the corresponding channel estimation error, respectively. The channel estimation errors are independent of the signal and the noise. Notice that matrix expectations keep the quadratic nature of the original QMP problems \cite{Gupta00} and therefore QMP still works for robust beamforming design \cite{XingICASSP2010}.

\section{Conclusions}

In this paper, we discussed linear transceiver designs with MSE as the performance criterion. Different from the previous existing works, the transceiver designs were understood from a unified optimization problem named QMP problems for various wireless systems.  The QMP based transceiver design algorithm discussed in this paper can be applied to a wide range of problems such as multi-cell coordination beamforming design, multi-user MIMO beamforming design, cognitive radio MIMO beamfoming design, cooperative network beamforming design and even their corresponding  robust designs with randomly distributed channel estimation errors. Meanwhile, the elegant properties of QMP problems were also discussed in detail in our work.

\appendices

\section{Monotonicity of $g(\mu)$ }
In order to prove that $g(\mu)$ is a monotonically decreasing function with respect to $\mu$, we assume that $\mu_1 \ge \mu_2$
 and in the following we will prove that $g(\mu_1) \le g(\mu_2)$.When $\mu_1 \ge \mu_2$, based on the fact that ${\bf{A}}$ is a positive semi-definite matrix we have \cite{Horn85} $({\bf{A}}_1^{-1/2}{\bf{A}}_0{\bf{A}}_1^{-1/2}+\mu_1{\bf{I}})\ge
({\bf{A}}_1^{-1/2}{\bf{A}}_0{\bf{A}}_1^{-1/2}+\mu_2{\bf{I}})$, based on which, taking inversion of both sides, the following inequality holds $({\bf{A}}_1^{-1/2}{\bf{A}}_0{\bf{A}}_1^{-1/2}+\mu_1{\bf{I}})^{-1}
\le({\bf{A}}_1^{-1/2}{\bf{A}}_0{\bf{A}}_1^{-1/2}+\mu_2{\bf{I}})^{-1}$. Therefore, it can be  concluded that
\begin{align}
&{\rm{Tr}}[{\bf{B}}_0{\bf{A}}_1^{-1/2}({\bf{A}}_1^{-1/2}{\bf{A}}_0{\bf{A}}_1^{-1/2}+\mu_1{\bf{I}})^{-2}
{\bf{A}}_1^{-1/2}{\bf{B}}_0^{\rm{H}}] \nonumber \\
& \le {\rm{Tr}}[{\bf{B}}_0{\bf{A}}_1^{-1/2}({\bf{A}}_1^{-1/2}{\bf{A}}_0{\bf{A}}_1^{-1/2}+\mu_2{\bf{I}})^{-2}
{\bf{A}}_1^{-1/2}{\bf{B}}_0^{\rm{H}}]. \nonumber
\end{align}

\end{document}